\documentclass[twocolumn,showpacs,preprintnumbers,amsmath,amssymb,showkeys]{revtex4}

\usepackage{amssymb}
\usepackage{amsmath}
\usepackage{bm}
\usepackage[latin1]{inputenc}
\usepackage[pctex32]{graphics}
\usepackage{graphicx}

\begin{document}
\newcommand{\be}{\begin{equation}}
\newcommand{\ee}{\end{equation}}
\newcommand{\ben}{\begin{eqnarray}}
\newcommand{\een}{\end{eqnarray}}
\newcommand{\nn}{\nonumber \\}
\newcommand{\ii}{\'{\i}}
\newcommand{\pp}{\prime}
\newcommand{\expq}{e_q}
\newcommand{\lnq}{\ln_q}
\newcommand{\quno}{q-1}
\newcommand{\qunoinv}{\frac{1}{q-1}}
\newcommand{\tr}{{\mathrm{Tr}}}
\newcommand{\nd}{\noindent}
\title{Semiclassical statistical mechanics' tools for deformed algebras}

\author{F. Olivares$^{1}$, F.~Pennini$^{1,2}$,  A.~Plastino$^2$, and G.L. Ferri$^3$ }

%\author{F.~Pennini}
%\thanks{E-mail:~pennini@fisica.unlp.edu.ar}

%\author{A.~Plastino}
%\thanks{E-mail:~plastino@fisica.unlp.edu.ar}

 \affiliation{$^1$Departamento de F\'{\i}sica, Universidad Cat\'olica del Norte,
 Av. Angamos 0610, Antofagasta, Chile\\
 $^2$Instituto de F\'{\i}sica La Plata
(CONICET--UNLP), Fac.\ de Ciencias Exactas, Universidad Nacional de
La Plata, C.C.~67, 1900 La Plata, Argentina\\$^3$ Facultad de
Ciencias Exactas, Universidad Nacional de La Pampa
 \\ Peru y Uruguay, Santa Rosa, La Pampa,
 Argentina}

%\date{\today}
\begin{abstract}
\nd In order to enlarge the present arsenal of semiclassical toools we explicitly obtain here the Husimi distributions and Wehrl
entropy within the context of
 deformed algebras built up on the basis of a new family of $q-$deformed
 coherent states, those of Quesne [{\it J. Phys. A} {\bf 35}, 9213
 (2002)].  We introduce also a generalization of the Wehrl entropy
 constructed with  escort distributions. The two generalizations are
 investigated with emphasis on i) their behavior as a function of temperature
 and ii) the results obtained when the deformation-parameter tends to unity.
\end{abstract}
%%05.20.-y: Classical statistical mechanics
%%05.40.-a: Fluctuation fenomena, ramdom process, noise....
%%05.70.-a: Thermodynamics
\maketitle

\section{Introduction}
\nd The semiclassical  approach  has had a long and
distinguished history and is a  very important weapon in the
physics' armory. Indeed, semiclassical approximations to quantum
mechanics remain an indispensable tool in many areas of physics
and chemistry. Despite the extraordinary evolution of computer
technology in the last years, exact numerical solution of the
Schrödinger equation is still quite difficult for problems with
more than a few degrees of freedom. Another great advantage of the
semiclassical approximation lies in that it facilitates an
intuitive understanding of the underlying physics, which is
usually hidden in blind numerical solutions of the Schrödinger
equation. Although semiclassical mechanics is as old as the
quantum theory itself, the field  is continuously evolving. There
still exist many open problems in the mathematical aspects of the
approximation as well as in the quest for new effective ways to
apply the approximation to various physical systems (see, for
instance, \cite{book1,book2} and references therein).

\vskip 2mm

\nd In a different vein, applications of the so-called
$q$-calculus to
 statistical mechanics have accrued increasing interest
 lately \cite{gellmann}. This $q$-calculus \cite{borges} has its origin in the
$q$-deformed harmonic oscillator theory, which, in turn, is based on
the construction of a $SUq(2)$ algebra of $q$-deformed commutation
or anti-commutation relations between creation and annihilation
 operators \cite{q1,q2,q3}. The above mentioned applications
 also employ ``deformed information measures" (DIM) that have been
  applied to different scientific
disciplines (see, for example, \cite{euro,t_jsp52,gellmann} and
references therein). DIMs were introduced long ago in the
cybernetic-information communities by Harvda-Charvat \cite{l1} and
Vadja  \cite{l2} in 1967-68, being
  rediscovered by Daroczy in 1970  \cite{l3}  with several echoes mostly in
the field of image processing. For a historic summary and the
pertinent references see Ref. \cite{okamoto}. In astronomy,
physics, economics, biology, etc., these deformed   information
measures are often called $q$-entropies since~1988~\cite{t_jsp52}.
\vskip 2mm

\nd In this work  we are concerned with  {\sf semiclassical
statistical physics'} problems and  {\it wish to add tools to the semiclassical armory} by, in particular,
investigating putative deformed {\it extensions} of two of its most
important quantities, namely,
 Husimi distributions (HD) and Wehrl entropies (WE). We will work within the context of
 deformed algebras built up with the new family of $q$-deformed
 coherent states introduced recently by Quesne \cite{Quesne},
 analyzing the  main HD and WE properties. \vskip 2mm

\nd Our attention will be focused on the  thermal
  description of the harmonic oscillator (HO) (together with its
  phase-space delocalization as temperature grows), in the understanding that {\sf
  the HO is, of course,  much more than a mere example},
  since in addition to the extensive use Glauber states in molecular physics and chemistry
   \cite{Klauder,molecular},  nowadays the HO
is of particular interest for the dynamics of bosonic or fermionic
atoms contained in magnetic traps \cite{davis,bradley}, as well as
for any system that exhibits an equidistant level spacing in the
vicinity of the ground state, like nuclei or Luttinger liquids.

\vskip 2mm \nd The paper is organized as follows: in Section \ref{semiclassicalHD} we
introduce the behavior of the HD and the WE for the HO, studying
the fluctuations in thermal equilibrium while, in Section \ref{QuesneCS},  the
new $q$-deformed coherent states are introduced. We study the
behavior of our generalizations of both  the HD and the WE in
Sections \ref{HusimiHD} and \ref{WehrlDE}, respectively. Generalized WE's bounds are
analyzed in Section \ref{Wehrlbound} while, in Section \ref{escort}, we deal with escort
distributions, which allow us to build up an alternative
generalization of the WE. Finally, some conclusions are drawn in
Section \ref{concluding}.

%%%%%%%%%%%%%%%%%%%%%%%%%%%%%%%%%%%%%%%%%%%%%
\section{Semiclassical distribution in phase-space}
%%%%%%%%%%%%%%%%%%%%%%%%%%%%%%%%%%%%%%%%%%%%%%
\label{semiclassicalHD}
\nd Wehrl's entropy $W$ is a very useful measure of localization in
phase-space~\cite{Wehrl}. It is  built up using coherent states
\cite{Glauber,PRD2753_93,Wehrl} and
  constitutes a
 powerful
 tool in statistical physics. The pertinent definition reads

\be W=-\int \frac{\mathrm{d}x\, \mathrm{d}p}{2 \pi \hbar}\,
 \mu(x,p)\, \ln \mu(x,p),\label{i1}\ee
      where $\mu(x,p)=\langle z| \rho|z\rangle$
   is a ``semi-classical'' phase-space distribution function
 associated to the density matrix $\rho$ of the
 system~\cite{Glauber,Klauder}.  Coherent states are eigenstates of the
 annihilation operator $a$, i.e., satisfies $a \vert z \rangle=z\vert z \rangle$.

  \nd The
 distribution  $\mu(x,p)$ is
   normalized in the fashion
   \be \int (\mathrm{d}x\, \mathrm{d}p/2 \pi \hbar)\,\mu(x,p)=1, \ee
   and it is often referred to as the Husimi
  distribution~\cite{Husimi}.   The last two equations clearly  indicate that the Wehrl
entropy is simply the ``classical entropy" ({\ref{i1}}) of a
Wigner-distribution. Indeed,   $\mu(x,p)$ is a
 Wigner-distribution $D_W$ smeared over an $\hbar$ sized region of phase space
 \cite{PRD2753_93}.
   The smearing renders $\mu(x,p)$ a positive function, even if $D_W$
  does not have
    such a character. The semi-classical Husimi probability
distribution
 refers to a special type of probability:
  that for simultaneous but approximate location of position and
 momentum in phase space~\cite{PRD2753_93}.
The uncertainty principle manifests itself  through the inequality $
\label{UPHO} 1 \leq W $, which was first conjectured by
Wehrl~\cite{Wehrl} and later proved by Lieb~\cite{Lieb}.

 \nd  The usual treatment of equilibrium in statistical mechanics makes
 use of
 the Gibbs's canonical distribution,  whose
    associated, ``thermal''
     density matrix is given by  \be \rho=Z^{-1}e^{-\beta H},\ee with $Z=\tr(e^{-\beta H})$  the partition function,
 $\beta=1/k_BT$
     the inverse  temperature $T$,
    and $k_B$ the Boltzmann constant.
In order to conveniently write down an expression for
    $W$ consider an arbitrary Hamiltonian $H$ of eigen-energies $E_n$
     and eigenstates   $|n\rangle$ ($n$ stands for a collection of all the pertinent
     quantum numbers required to label the states).
 One can
  always write~\cite{PRD2753_93}
 \be \mu(x,p)=\frac{1}{Z} \sum_{n}
 e^{-\beta
   E_n}|\langle z|n\rangle|^2.\label{husimi0}
 \ee A useful route to $W$ starts then with Eq.~(\ref{husimi0}) and
 continues with Eq.~(\ref{i1}). In the special case of the harmonic oscillator the coherent states are of the form~\cite{Glauber}

 \be
 |z\rangle=e^{-|z|^2/2}\,\sum^{\infty}_{n=0}\frac{z^{n}}{\sqrt{n!}}\,|
    n\rangle, \label{qcoherent}
 \ee
where ${|n\rangle}$ are a complete orthonormal set of eigenstates
and whose spectrum of energy is $E_n=(n+1/2)\hbar\omega$,
$n=0,1,\ldots$ In this situation the analytic expression for the HD
and the WE were obtained in Ref.~\cite{PRD2753_93}

\be \mu(z)=(1-e^{-\beta\hbar\omega})\,e^{-(1-e^{-\beta\hbar\omega
})|z|^2}\label{mu1},
 \ee
\begin{equation}\label{wehrl}
    W=1-\ln(1-e^{-\beta\hbar\omega}).
\end{equation}
When $T\rightarrow 0$, the entropy takes its minimum value $W=1$,
expressing purely quantum fluctuations. On the other hand when
$T\rightarrow \infty$, the entropy tends to the value
$-\ln(\beta\hbar\omega)$ which expresses purely thermal
fluctuations.

%%%%%%%%%%%%%%%%%%%%%%%%%%%%%%%%%%%%%%%%%%%%%%%%%%%%%%%%%%
\section{Quesne's new $q$-deformed coherent states}
%%%%%%%%%%%%%%%%%%%%%%%%%%%%%%%%%%%%%%%%%%%%%%%%%%%%%%%%%%
\label{QuesneCS}
\nd  Quesne advanced in Ref. \cite{Quesne} a new family of
harmonic oscillator physical states, labelled by $0<q<1$ and $z\in~\mathbb{C}$,

\begin{equation}
    |z\rangle_{q}=N_{q}^{-1/2}\,\sum^{\infty}_{n=0}\frac{z^{n}}{\sqrt{[n]_{q}!}}\,|
    n\rangle, \label{2qcoherent}
\end{equation}
 where $|n\rangle=(n!)^{-1/2}(a^{\dagger})^{n}|0\rangle$ is the $n$-boson
state and one introduces quasi-factorials \cite{Quesne}
\begin{displaymath}
    [n]_{q}!\equiv\left\{ \begin{array}{ll}
  [n]_{q}[n-1]_{q} \ldots [1]_{q} \quad \quad & \textrm{if}\, \,\, n=1,2,3 \ldots\\
  1\quad \quad &\textrm{if} \,\,\, n=0
\end{array}\right.
\end{displaymath}
with
\begin{equation}
    [n]_{q}\equiv\frac{1-q^{-n}}{q-1}=q^{-n}\{n\}_{q}. \label{qfactorial}
\end{equation}
$\{n\}_{q}$ is called the ``$q$-basic number" and generates its
own factorial  (the $q$-factorial) $\{n\}_{q}!$ \cite{Quesne}. The
$q$-factorial
 can also be written in terms of the $q$-gamma function~$\Gamma_{q}(x)$~\cite{Quesne}

\begin{equation}
    [n]_{q}!=q^{\frac{-n(n+1)}{2}}\,\Gamma_{q}(n+1)=q^{\frac{-n(n+1)}{2}}\{n\}_{q}!,
\end{equation}
that, in the limit $q\rightarrow1$, yields $[n]_{q}$, $[n]_{q}!$,
and $\Gamma_{q}(n)$ tending  to $n$, $n!$, and $\Gamma(n)$,
respectively. The states $|z\rangle_{q}$ allow one to build up
$q$-deformed coherent states. They will be acceptable generalized
coherent states if  three basic mathematical properties are verified
\cite{Klauder}. The states $|z\rangle_{q}$ should be: i)
normalizable, ii) continuous in the $z$-label. Additionally,
iii)~one must ascertain  the existence of a resolution of unity with
a positive weight function. The normalization condition,
$_{_{q}}\langle z|z\rangle_{q}=1$ leads to
\begin{equation}
    N_{q}(|z|^{2})=\sum^{\infty}_{n=0}\frac{(|z|^{2})^{n}}{[n]_{q}!}=E_{q}[(1-q)q|z|^{2}],
\end{equation}
where $E_{q}(x)\equiv\prod^{\infty}_{k=0}(1+q^{k}x)$ is one of the
so-called Jackson's $q$-exponentials introduced in 1909 \cite{Jackson} such that
$\lim_{q\rightarrow1}E_{q}[(1-q)x]=e^{x}$. Since $N_{q}(|z|^{2})$
is equal to $E_{q}[(1-q)q|z|^{2}]$ (a well defined function in
$0<q<1$), the new  states are normalizable on the whole complex
plane. On the other hand, the states $|z\rangle_{q}$ are always
continuous in~$z$. Finally, for the resolution of unity one needs

\begin{equation}
    \int\int_{c}d^{2}z|z\rangle_{q}K_{q}(|z|^{2}) _{q}\langle
    z|=\sum^{\infty}_{n=0}|n\rangle\langle n|=I,
\end{equation}
with a weight function $K_{q}(|z|^{2})$ that can be obtained using
 the expressions for $|z\rangle_{q}$ and its conjugate and then
performing  the integral by recourse to the $q$-analogue of the
Euler gamma integral \cite{Quesne}. Quesne finds

\begin{equation}
    K_{q}(|z|^{2})=\frac{1-q}{\pi\ln q^{-1}}\frac{E_{q}[(1-q)q|z|^{2}]}{E_{q}[(1-q)|z|^{2}]}.
\end{equation}
It is seen that, in the limit $q\rightarrow1^{-}$, we have $
K_{q}(|z|^{2})\rightarrow K(|z|^{2})=1/ \pi$, corresponding to the
weights for the conventional coherent states of the harmonic
oscillator. This entails that we indeed have at hand new
$q$-deformed coherent states that fulfill the standard properties.
In the next section we start presenting the  results of this
communication, using these Quesne-coherent states to generalize our
$q$-Husimi distribution.

%%%%%%%%%%%%%%%%%%%%%%%%%%%%%%%%%%%%%%%%%%%%%%%%%%%%
\section{The $q$-Husimi distribution to $0<q<1$}
%%%%%%%%%%%%%%%%%%%%%%%%%%%%%%%%%%%%%%%%%%%%%%%%%%%%%
\label{HusimiHD}

\nd We define the $q$-Husimi distribution ($q$-HD) in the rather
``natural" fashion

\begin{equation}\label{qh}
    \mu_{q}(z)\equiv\, _{q}\langle z|\rho|z\rangle_{q},
\end{equation}
using Quesne's HO-$q$-coherent states (\ref{qcoherent}). From these
it is easy to find  an analytic expression for our new $q$-Husimi
distributions

\begin{equation}\label{qhusimi}
    \mu_{q}(z)=
    (1-e^{-\beta\hbar\omega})
    \,\frac{E_{q}[(1-q)q|z|^{2}e^{-\beta\hbar\omega}]}{E_{q}[(1-q)q|z|^{2}]}.
\end{equation}
 It is important to remark that

\begin{equation}\label{qlim}
    \lim_{q\rightarrow1}\mu_{q}(z)=\mu(z)
\end{equation}
where $\mu(z)$ is given by (\ref{mu1}). We have  numerically
ascertained that our $q$-Husimi distribution is normalized using
the same weight function $K_{q}(|z|^{2})$ that Quesne employed for
the resolution of unity of her $q$-coherent states.

\begin{equation}\label{qnorhusimi}
    \int_{c}d^{2}zK_{q}(|z|^{2})\mu_{q}(z)=1.
\end{equation}

\begin{figure}
\begin{center}
%\hspace{0cm}
\includegraphics[width=7cm]{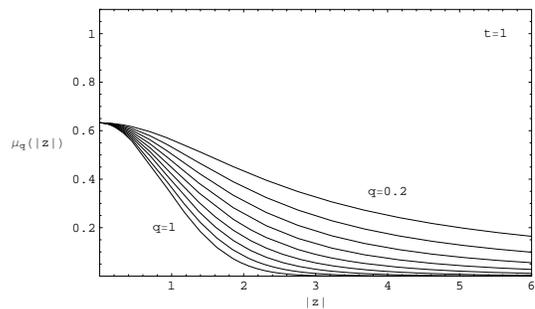}\\
 \caption{\footnotesize\label{q-husimi} {\it{$\mu_{q}(z)$ vs. $|z|$ for two
 different normalized temperatures  $t=T/\hbar\omega$, and several values of $q$.}}}
 \label{husimifig}
\end{center}
\end{figure}

\nd We see in  Fig. \ref{husimifig} that a  $q$-HD's height depends
only upon the temperature (not on $q$!).  $q$-Deformation affects
only shapes.  Thus,
\begin{equation}\label{10}
    0<\mu_{q}(z)\leq1,
\end{equation}
and $\mu_{q}(z)$ remains a legitimate semi-classical probability
distribution (unlike Wigner's one).

%%%%%%%%%%%%%%%%%%%%%%%%%%%%%%%%%%%%%
\section{Deformed Wehrl entropies}
%%%%%%%%%%%%%%%%%%%%%%%%%%%%%%%%%%%%%%
\label{WehrlDE}

\nd In order to introduce deformed Wehrl entropies we start by
reminding the reader of the concept of  $q$-deformed measures,
that read \cite{naudts}

 \be
 S_q=-k_B\,\int \mathrm{d}\Omega \,f(x)^q\, \ln_{q}
 f(x),
 \label{Sq}
\ee where $k_B$ is the constant of Boltzmann's,  $f(x)$ a
normalized probability density, and one defines the $q$-logarithm
function in the fashion \cite{naudts}
 \be \label{naudts} \ln_q\,x=(1-x^{1-q})/(q-1)\,\,\,
  {\rm with} \,~x>~0; \,\,\,q \in \Re, \ee where $q$
is called the deformation index (the signature of the ``deformed"
nature of the  measure). For $q=1$ we reobtain the ordinary
logarithm and the logarithmic Shannon measure. The measure
(\ref{Sq}) can also be generated using the $q-$calculus introduced
by Jackson \cite{Jackson} in the following form \cite{Abe_q} \be
S_q=-k_B D_{\alpha}^{(q)} g( \alpha)\big|_{\alpha=1}, \ee where
$g(\alpha)=\int \mathrm{d}x f(x)^{\alpha}$ and $D^{(q)}_x$ is
Jackson's $q$-derivative operator  \be D^{(q)}_x=\frac{h(q
x)-h(x)}{q x-x}, \ee which reduces to the ordinary Leibnitz-one
$d/dx$ when the parameter $q$ goes to unity~\cite{Abe_q}. \vskip
2mm

\begin{figure}[h]
\centering
\includegraphics[width=8cm]{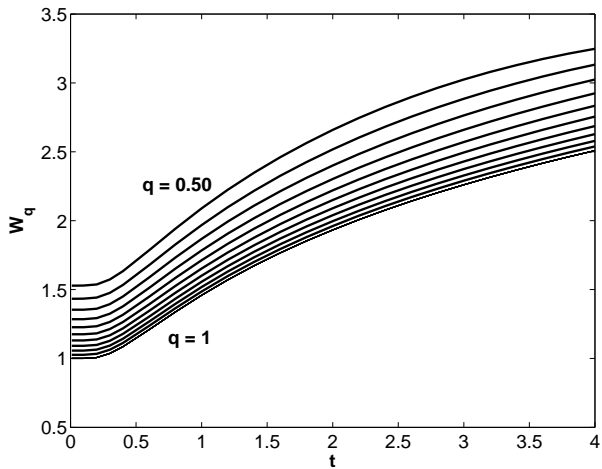}
 \caption{ $W_{q}(\mu_{q})$ vs. $t$ ($t=T/\hbar\omega$), for values ranging
 downwards from $q=0.95$
  $q=0.50$. The standard standard WE ($q=1$) is also represented.}
  \label{qWehrl}
\end{figure}

\nd Accordingly, the natural definition of  a deformed Wehrl
entropy reads as  follows
\begin{equation}\label{qw}
    W_{q}\equiv -\int_{c}d^{2}zK_{q}(|z|^{2})\mu_{q}(z)\ln \mu_{q}(z),
\end{equation}
and, inserting  the explicit HO-form (\ref{qhusimi}) into the above
expression,  we find

\begin{equation}\label{qwehrl2}
    W_{q}=f(q)-\ln(1-e^{-\beta\hbar\omega})\equiv f(q)-1 +W,
\end{equation}
where
\begin{equation}\label{fq}
    f(q)=-\int
    K_{q}\,d^{2}z\,\mu_{q}\ln\left\{\frac{\mu_{q}}{1-e^{-\beta\hbar\omega}}\right\}.
\end{equation}
It easy to check that in the limit $q\rightarrow1$ one has $f(q)
\rightarrow1$ and Eq. (\ref{qwehrl2}) leads to  the usual Wehrl
entropy. The $q$-Wehrl entropy can not be obtained in analytic
fashion and needs numerical evaluation. In Fig. \ref{qWehrl} we
 plot  $W_{q}$ vs. $t$ ($t=T/\hbar\omega$) for several values of the parameter~$q$.

We see from (\ref{fq}) and comparison with the conventional HO-Wehrl
entropy $W=1-\ln(1-e^{-\beta\hbar\omega})$ that  deformation merely
entails a constant-in-$z$ additive function $f(q)$, depending solely
on $q$, as depicted in Fig. \ref{qWehrl}, so that $W_q$ behaves,
as a function of deformation, exactly like the $q$-distributions of
the previous Sections. At low enough temperatures our $q$-coherent
states minimize the $q$-entropy, i.e.,  $W_{q}\sim1$, yielding
maximal localization. Delocalization grows with $T$, of course.

\vskip 3mm \nd At this stage we are in  a position to draw an
important consequence. In general, deformed (or Tsallis')
entropies are quite different objects as compared with Shannon's
one, save for $q$ close to unity. This is not the case for Wehrl
entropies! $q$-deformation becomes in this case  a smooth
shape-deformation. At the semiclassical level the difference
between Tsallis and Shannon entropies becomes weaker than either
at the quantal or the classical levels.

%%%%%%%%%%%%%%%%%%%%%%%%%%%%%%%%%%%%%%%
\section{Wehrl entropy bounds}
%%%%%%%%%%%%%%%%%%%%%%%%%%%%%%%%%%%%%%%%%
\label{Wehrlbound}

\nd Using the definition of the $q$-WE given by Eq.
(\ref{qwehrl2}) we can easily investigate  possible bounds for the
Wehrl entropy, i.e., for our localization power. We  see that,
when $T\rightarrow0$, one has
\begin{equation}\label{Wl1}
    W_{q}\rightarrow \mathcal{A}_q\equiv \frac{(q-1)}{\ln(q)}\,g(q),
\end{equation}
where
\begin{equation}\label{g}
    g(q)=\int^{\infty}_{0}d|z|^{2}
    \frac{\ln(E_{q}[(1-q)q|z|^{2}])}{E_{q}[(1-q)|z|^{2}]},
\end{equation}
while, for $T\rightarrow\infty$,
\begin{equation}\label{wl2}
    W_{q}\rightarrow -\ln(\beta\hbar\omega)=\infty,
\end{equation}
having a lower bound only. Since, obviously,  $W_q$ is a
monotonously growing function of $T$ we can state that
\begin{equation}\label{Wl11}
    \mathcal{A}_q\leq W_{q},
\end{equation}
which constitutes the new deformed Lieb-relation. It is easy to
check the limit $q\rightarrow1$:
\begin{equation}\label{lim}
    \lim_{q\rightarrow1}\mathcal{A}_q=1,
\end{equation}
and we reobtain the known Lieb bound of the standard Wehrl entropy
$1\leq W$.  In
Table \ref{list1} we illustrate the behavior of the function $\mathcal{A}_q$
vs. $q$. We note that the function reproduces the values of the
$q$-Wehrl entropy at $T=0$, as expected.
\begin{table}[h]
\begin{center}
\caption{$\mathcal{A}_q$ vs. $q$}
\label{list1}
\begin{tabular}{lrrccl}
  \hline
  $q$ & $\mathcal{A}_q$ \\
  \hline
  \hline
  0.50 & 1.5289 \\
  0.55 & 1.4337 \\
  0.60 & 1.3536 \\
  0.65 & 1.2852 \\
  0.70 & 1.2264 \\
  0.75 & 1.1753 \\
  0.80 & 1.1307 \\
  0.85 & 1.0916 \\
  \hline
\end{tabular}
\end{center}
\end{table}

%%%%%%%%%%%%%%%%%%%%%%%%%%%%%%%%%%%%
\section{Escort generalizations}
%%%%%%%%%%%%%%%%%%%%%%%%%%%%%%%%%%%%
\label{escort}
\nd In this Section we consider a new possible deformation of the
entropy, which we shall call $q$-escort Wehrl entropy ($q$-eWE).
It is build up from a deformation of the HD advanced in Ref.
\cite{pennini}, called the $q$-escort Husimi distribution
($q$-eHD). In general, given a normalized probability density (PD)
$f(x)$, its associated q-escort PD $F(x)$ reads \be F(x)=
\frac{f(x)^q}{\int\,dx\,f(x)^q}. \ee \nd If we replace here $f(x)$
by the Husimi distribution one gets \cite{pennini}
 \begin{equation}\label{qhusimi
e2}
    \gamma_{q}(|z|)=q(1-e^{-\beta\hbar\omega}) \,
    \exp[-q|z|^{2}(1-e^{-\beta\hbar\omega})],
\end{equation}
where $q$ $\in$ $(1,\sqrt{2})$ \cite{pennini}.

\nd Our goal is now to get the associated $q$-Wehrl's entropy.
 Replacing Eq. (\ref{qhusimi e2}) into Eq. (\ref{i1}), we immediately
find
\begin{equation}
    W_{(q)}(\gamma_{q})=q^{2}(1-e^{-\beta\hbar\omega})^{2}I_{q}-\ln[q(1-e^{-\beta\hbar\omega})]
    N_{q},
\end{equation}
where
\begin{equation}
 I_{q}=\int\frac{d^{2}z}{\pi}|z|^{2}
 \,\exp[-q|z|^{2}(1-e^{-\beta\hbar\omega})],
\end{equation}
and
\begin{equation}\label{ne}
    N_{q}=\int\frac{d^{2}z}{\pi}\gamma_{q}(|z|).
\end{equation}
Integrating over all phase space one finds $N_{q}=1$,  the
normalization condition of the $q$-eHD. On the other hand we have
$I_{q}=1/q^{2}(1-e^{-\beta\hbar\omega})^{2}$. Thus, our $q$-eWE,
is given by
\begin{equation}\label{qw3}
    W_{(q)}(\gamma_{q})=1-\ln[q(1-e^{-\beta\hbar\omega})]=W-\ln(q),
\end{equation}
a rather interesting result that entails, for our alternative
escort $q$-deformation, that it simply  adds a new  term to the
entropy, given by $\ln (q)$ (vanishing, as it should, for $q=1$).

\begin{figure}
\begin{center}
%\hspace{0cm}
\includegraphics[width=7cm]{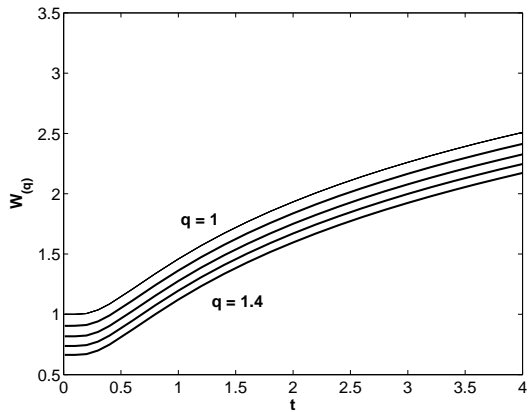}
%\vspace{0cm}
 \caption{\footnotesize\label{qwehrle} {\it{$q$-EWe vs. $t$ ($t=T/\hbar\omega$),
 for $q=1.1$ to $q=1.4$. The standard WE ($q=1$) is also represented}}}
\end{center}
\end{figure}

\nd Analyzing the behavior of the $q$-eWE with temperature we can
look for its  bounds and  easily ascertain that
\begin{eqnarray*}
  \nonumber T\rightarrow0 &\Rightarrow& W_{(q)}(\gamma_{q})\rightarrow1-\ln(q) \\
  \nonumber T\rightarrow\infty &\Rightarrow& W_{(q)}(\gamma_{q})\rightarrow
  -\ln(\beta\hbar\omega)=\infty,
\end{eqnarray*}
i.e., only a  lower bound exists. In table \ref{tab22} we
illustrate how the lower bound changes with $q$.  Fig. \ref{qwehrle} depicts $W_{(q)}$ vs. $t$ for several values of
$q$. The $q$-escort entropies is able to ``perforate" the Lieb
lower bound for $q \ge 1$, without violating the uncertainty
principle, since  $q \in (0,\sqrt{2})$.

\begin{table}[ht]
%{\bf Tabla~\ref{sources.tab}.}
\begin{center}
\caption[]{{\it{$1-\ln(q)$ for different values of $q$.}}}
\label{tab22}
\begin{tabular}{lrrccl}
\multicolumn{2}{c}{}\\
\multicolumn{1}{l}{$q$} & \multicolumn{1}{c}{$1-\ln(q)$} &
\multicolumn{2}{c}{} \\

\hline \hline
1.1    & 0.904690\\
1.2    & 0.817678\\
1.3    & 0.737636\\
1.4    & 0.663528\\
\hline
\end{tabular}
\end{center}
\end{table}

%%%%%%%%%%%%%%%%%%%%%%%%%%%%%%%%%%%%%%%%%%
%\newpage
%%%%%%%%%%%%%%%%%%%%%%%%%%%%%%%%%%%%%%%%%%
\section{Conclusions}
%%%%%%%%%%%%%%%%%%%%%%%%%%%%%%%%%%
\label{concluding}

Semiclassical Husimi distributions and their
associated Wehrl entropies have been here investigated within the  frame of deformed algebras.
As a summary of our results we can state that: \begin{itemize}

\item  we have advanced a $q$-generalization of the Husimi distribution
$\mu_{q}(z)$, which arises from the Quesne-family of $q$-coherent
states and found that the $q$-deformation does not change the HD's
property of being legitimate probability distributions, i.e.,
$0<\mu_{q}\leq1$.

\item  the above leads to a concomitant generalization of the WE $W_{q}(\mu_{q})$,
whose lower bound coincides with  the well-known Lieb one. These
semiclassical $q$-entropies approach the standard one when $q$
tends to unity.

\item  a different generalization of the Wehrl entropy,  $W_{(q)}(\gamma_{q})$, called
the escort-one has also been advanced in Section \ref{escort}, starting
from the $q$-eHD of Ref. \cite{pennini}. We saw that this
generalization closely resembles
 the standard WE, but its lower bound improves on
 the  Lieb one, allowing (at least in principle) for a better localization in
phase-space. This point  should be further considered in future research. \end{itemize}

%%%%%%%%%%%%%%%%%%%%%%%%%%%%%%%%%%%%%%%%%

\section{Acknowledgments}

F. Olivares and F. Pennini would like to thank partial financial
support by DGIP-UCN 2006.


\begin{thebibliography}{99}

\section{References}

\bibitem{book1} M. Dimassi and J. Sjoestrand, {\it Spectral Asymptotics in the
Semi-Classical Limit}, (Cambridge University Press, Cambridge, UK,
1999).

\bibitem{book2} M. Brack and R. K. Bhaduri, {\it Semiclassical Physics},
(Addison-Wesley, Reading, MA, 1997).



\bibitem{gellmann}   M.~Gell-Mann and C.~Tsallis, Eds.
{\it Nonextensive Entropy: Interdisciplinary applications} (Oxford
University Press, Oxford, 2004), and references therein %1

\bibitem{borges} E.P. Borges, {\it Physica A} {\bf 340} (2004) 95. %%%2

\bibitem{q1} L. Biedenharn, J. Phys. A 22, L873 (1989) %3
\bibitem{q2} A. Macfarlane, J. Phys. A 22, 4581 (1989) %4
\bibitem{q3} J.L. Gruver, Phys. Lett. A 254, 1 (1999) %5

\bibitem{t_jsp52} C.~Tsallis, J.~Stat.~Phys. \textbf{52} (1988) 479 %%6


\bibitem{euro} J.P. Boon and C. Tsallis, eds., Nonextensive Statistical
Mechanics: New Trends, new perspectives, Europhysics
News {\bf 36}, Number 6 (2005);
 M.~Buchanan, {\it New Scientist}
\textbf{187}, (2005) 34. %%7


%%\bibitem{gellmann}   M.~Gell-Mann and C.~Tsallis, Eds.
%%%%%%%%{\it Nonextensive Entropy: Interdisciplinary applications}
%%(Oxford
%%University Press, Oxford, 2004), and references therein %16


\bibitem{l1} J. Havrda and F.  Charvát, ``Quantification method of classification
processes. Concept of structural $a$-entropy." 1967 Kybernetika
(Prague) {\bf 3}  30 %%8


\bibitem{l2} I. Vajda, ``Igor Axioms for $a$-entropy of a generalized probability
scheme." 1968 (Czech) Kybernetika (Prague) {\bf 4} 105 %9


\bibitem{l3} Z. Daróczy, ``Generalized information functions." 1970 Information and Control
{\bf 16} 36 %%10

\bibitem{okamoto} S.~Abe and Y.~Okamoto, Eds. {\it Nonextensive
statistical mechanics and its applications} (Springer Verlag,
Berlin, 2001) %%11



\bibitem{Quesne} C. Quesne, {\it J. Phys. A} {\bf 35}, 9213 (2002) %%15

%%%%%%%%%%%%%%%%%%
\bibitem{Klauder} J. R. Klauder and B.-S. Skagerstam,
{\it Coherent States, Applications in Physics and Mathematical
Physics} (World Scientific, Singapore, 1985)

\bibitem{molecular}
 E. Deumens, A. Diz, R. Longo, and Y. Oehrn, ``Time-dependent
theoretical treatments of the dynamics of electrons and nuclei in
molecular systems'', {\bf 66}, 917 (1994)
%%''Time-dependent theoretical treatments of the dynamics of electrons and nuclei in
% molecular systems''(Rev. Mod. Phys.)


\bibitem{davis} K.B.~Davis {\it et al.}, ``Bose-Einstein Condensation in a Gas of Sodium Atoms'',
{\it Phys. Rev. Lett.} {\bf 75}, 3969 (1995)
%%''Bose-Einstein Condensation in a Gas of Sodium Atoms''

\bibitem{bradley} C.C.~Bradley, C.A.~Sackett, and R.G.~Hulet,
``Bose-Einstein Condensation of Lithium: Observation of Limited
Condensate Number''
 {\it Phys. Rev. Lett.} {\bf 78}, 985 (1997)
%%''Bose-Einstein Condensation of Lithium: Observation of Limited Condensate Number''

%%%%%%%%%%%%%%%%%


\bibitem{Wehrl} A.~Wehrl, Rep.~Math.~Phys. \textbf{16} (1979) 353 %%16
%%''General properties of entropy''.

%%%%%%%%%%%%%%%%%


\bibitem{Glauber} R.J.~Glauber,  Phys. Rev. {\bf 131} (1963) 2766 %%17
%% ''Coherent and Incoherent States of the Radiation field''.



\bibitem{PRD2753_93} A.~Anderson and J.J.~Halliwell,
Phys.~Rev.~D \textbf{48} (1993) 2753 %%19
 %%''Information-theoretic measure of uncertainty due to quantum and thermal
%%  fluctuations''


\bibitem{Husimi} K. Husimi, Proc. Phys. Math. Soc. Jpn. {\bf 22},
264 (1940) %% 20

\bibitem{Lieb} E.H.~Lieb, Commun.~Math.~Phys. \textbf{62} (1978) 35.
%%21

\bibitem{pennini} F. Pennini, A. Plastino and G. Ferri,
%``Semiclassical information from deformed and escort information
%measures."
Physica A {\bf{383}}, 2, 782-796 (2007). %%22


\bibitem{naudts} J. Naudts, {\it Physica A} {\bf 316} (2002) 323. %%12

\bibitem{Jackson} F.H. Jackson, {\it Q.J. Pure Appl. Math.} {\bf 41}, 193
(1910) %13



\bibitem{Abe_q} S. Abe, {\it Phys. Lett. A} 224 (1997) 326. %%14

\end{thebibliography}
\end{document}